\documentclass[conference]{IEEEtran}
\usepackage{cite}
\usepackage{amsmath,amssymb,amsfonts}
\usepackage{algorithmic}
\usepackage{graphicx}
\usepackage{textcomp}
\usepackage{xcolor}
\usepackage{orcidlink}
\usepackage[caption=false,font=footnotesize]{subfig}
\usepackage{listings}

\def\BibTeX{{\rm B\kern-.05em{\sc i\kern-.025em b}\kern-.08em
    T\kern-.1667em\lower.7ex\hbox{E}\kern-.125emX}}
\begin{document}

\title{Project-Based Learning in Introductory Quantum Computing Courses: A 
Case Study on Quantum Algorithms for Medical Imaging}

\author{
\IEEEauthorblockN{1\textsuperscript{st} Nischal Binod Gautam}
\IEEEauthorblockA{\textit{Department of Electrical and Computer Engineering} \\
\textit{Baylor University}\\
Waco, USA \\
Nischal\_Gautam1@baylor.edu \orcidlink{0009-0001-1230-9967}}
\and
\IEEEauthorblockN{2\textsuperscript{nd} Keith Evan Schubert}
\IEEEauthorblockA{\textit{Department of Electrical and Computer Engineering} \\
\textit{Baylor University}\\
Waco, USA \\
Keith\_Schubert@baylor.edu \orcidlink{0000-0003-2718-5087}}
\and
\IEEEauthorblockN{3\textsuperscript{rd} Enrique P. Blair}
\IEEEauthorblockA{\textit{Department of Electrical and Computer Engineering} \\
\textit{Baylor University}\\
Waco, USA \\
Enrique\_Blair@baylor.edu \orcidlink{0000-0001-5872-4819}}
}

\maketitle

\begin{abstract}
    Quantum computing introduces abstract concepts and non-intuitive 
    behaviors that can be challenging for students to grasp through traditional 
    lecture-based instruction alone. This paper demonstrates how Project-Based Learning (PBL) 
    can be leveraged to bridge that gap. This can be done by engaging students in a real-world, 
    interdisciplinary 
    task that combines quantum computing with their field of interest. 

    As part of a similar assignment, we investigated the application of the 
    Harrow-Hassidim-Lloyd (HHL) algorithm for computed tomography (CT) image reconstruction 
    and benchmarked its performance against the classical Algebraic Reconstruction Technique (ART). 
    Through implementing and analyzing both methods on a 
    small-scale problem, we gained practical experience with quantum algorithms, 
    critically evaluated their limitations, and developed technical writing and research skills.
    
    The experience demonstrated that Project-Based Learning not 
    only enhances conceptual understanding but also 
    encourages students to engage deeply with emerging technologies through research, 
    implementation, and reflection.
    We recommend the integration of similar PBL modules in 
    introductory quantum computing courses. 
    The assignment also works better if students are required to write and submit a 
    conference-style paper, supported by mentorship from faculty across 
    the different fields.
    In such course interdisciplinary, real-world problems 
    can transform abstract theory into meaningful learning experiences and better prepare 
    students for future advancements in quantum technologies.
\end{abstract}
     

\section{Introduction}

\subsection{Project-Based Learning}

Project-based learning (PBL) \cite{dewey1928} \cite{hand2018} is an inquiry driven, student-centred 
approach that challenges learners to tackle authentic problems and produce concrete outcomes.  
It is on John Dewey's philosophy of “learning by doing” \cite{waks} and William Heard Kilpatrick's 
early "project method \cite{heard} which held that powerful learning 
grows from activities students genuinely care about \cite{Peterson2012}. Contemporary 
PBL rests on some constructivist principles \cite{Kokotsaki2016}:

\begin{enumerate}
  \item \textbf{Context-specific learning \cite{skalet}}: Knowledge is 
  constructed in real-world situations rather than through abstract drills. 
  \item \textbf{Active involvement \cite{johnson2013}}: Students lead the process, posing 
  driving questions, designing investigations, gathering and analyzing data, 
  and iterating on solutions.
  \item \textbf{Social construction of understanding \cite{Cocco2006}}: Learning emerges through 
  collaboration, discussion, and the shared refinement of ideas.
\end{enumerate}
  
This approach helps learners grasp complex concepts, cultivates positive 
attitudes toward the subject matter, and sharpens reasoning and critical-thinking skills.  
In practice, students move cyclically through problem identification, research, 
data collection, analysis, strategy development, and product creation.  
Each phase is tightly integrated with coursework and includes coordinated 
individual, group, and classroom activities aimed at fostering high-level 
thinking skills. \cite{sahin} \cite{sahin2}

For beginners without a background in quantum mechanics, quantum computing may seem notoriously 
abstract. Students must juggle linear-algebraic formalisms, 
complex amplitudes, and non-intuitive circuit behavior that can feel detached from 
concrete experience.  Recent research in quantum education shows that after traditional 
lecture-based instruction, many learners still cling to faulty “reasoning primitives” \cite{Singh2024} 
(e.g.\ assuming an $N$-qubit computer simply upgrades every factor of $N$ by $2^N$ as compared 
to a classical computer) and 
struggle to visualize the vast state space available to a superposition. 

We suggest that students in introductory courses on quantum computing may 
benefit from PBL assignments
that require both individual research and technical writing. This paper provides an 
example of such an assignment used in a medical imaging course. 
We, the authors, found this to be enriching and recommend it for use in introductory 
Quantum Computing courses.

\subsection{Objectives}

Guided by the challenges and opportunities outlined above, this 
study pursues a dual agenda: pedagogical and technical, framed by project-based learning:
\begin{enumerate}
    \item Carry out a PBL module that bridges quantum computing and an arbitrary research field.\\
          Develop a scaffolded learning sequence in which students can build and test
          a Quantum Algorithm or apply knowledge of Quantum Computing to a 
          task related to an arbitrary research field.

    \item Evaluate educational impact.\\
          Assess the success of the PBL task by noting down any changes in the students'
          \begin{enumerate}
              \item conceptual understanding of the specific quantum computing primitives,  
              \item confidence in mapping domain problems to quantum formulations, and  
              \item ability to critique classical versus quantum solvers.
          \end{enumerate}
\end{enumerate}

\section{Example of PBL Integration in a Course: Image Reconstruction in Medical Imaging}

In this paper, we present as a model project, an exploration of Quantum Computing concepts 
applied to image processing. Specifically, we focus on 
an image reconstruction technique based on a quantum approach developed by Harrow, Hassidim, and Loyd 
named the Harrow-Hassidim-Loyd (HHL) algorithm. \cite{harrow2009} 
We demonstrate image reconstruction based on the HHL technique and benchmark it 
against the Algebraic Reconstruction Technique (ART) \cite{ander1989} technique. (for a toy problem). 
These calculations reveal that Noisy Intermediate-Scale Quantum (NISQ) \cite{lau} era quantum computers 
are not yet 
capable of handling industrial scale image reconstruction 
problems. We discuss advances in technology required to make HHL a viable tool for MI applications.
Finally, we conclude by highlighting the value of such a PBL assignment in the 
context of introductory quantum computing course. 
The assignment also included a technical writing component requiring students to prepare a 
formal conference-style paper and submit it to a conference of their choice. 

\subsection{Introduction of the Project}
The Algebraic Reconstruction Technique (ART) \cite{ander1989} is a widely used iterative method in Computed 
Tomography (CT) \cite{brooks} for image reconstruction.
ART reconstructs images by solving a system of linear equations \cite{gordon} derived from X-ray projection 
measurements:
\begin{equation}
A \vec{x} = \vec{b}
\end{equation}
Here, \(x\) is the vector of unknown pixel or voxel intensities, \(b\) is the measured 
projection data, and \(A\) is the system matrix that encodes the geometry of the CT 
scanner and the imaging process. ART updates an initial image estimate by iteratively 
reducing the discrepancy between measured and simulated projections:
\begin{equation}
x^{(k+1)} = x^{(k)} + \lambda \frac{b_i - a_i x^{(k)}}{\|a_i\|^2} a_i^T
\end{equation}
where \(a_i\) is the \(i^\text{th}\) row of matrix \(A\), \(b_i\) is the corresponding 
measurement, and \(\lambda\) is a relaxation parameter that controls the convergence behavior.
ART is particularly effective in handling incomplete or noisy data. However, its iterative 
nature can be computationally intensive for large-scale imaging tasks. \cite{ander1989} \cite{zhang2021}

Quantum computers leverage the principles of quantum mechanics to process 
information in fundamentally different ways than classical computers. They also 
offer exponential speedups for specific computational problems. In 2009, Harrow, Hassidim, 
and Lloyd introduced the HHL algorithm. \cite{harrow2009} It is a quantum algorithm capable of 
solving linear systems with exponentially faster time complexity than classical methods.

HHL encodes the input vector \(\vec{b}\) into a quantum state using amplitude encoding. This 
allows a \(2^N \times 1\) vector to be represented using only \(N\) qubits. Compared to 
other quantum algorithms like Quantum Phase Estimation (QPE) \cite{moham} and the Variational Quantum 
Eigensolver (VQE) \cite{tilly}, HHL achieves an exponential reduction in qubit requirements. 
This efficiency may
make it attractive for large-scale applications such as medical imaging.
\cite{harrow2009}

The HHL algorithm presents a distinct and promising possibility of replacing  ART in computed 
tomography. It has the ability to process large-scale linear systems efficiently. 
As quantum hardware improves, this opens the door 
to integrating HHL into real-world application for CT image reconstruction.
This will also result in reducing computation time and improving 
the handling of high-resolution imaging data. While ART remains the current standard, HHL 
is presented as a candidate with the potential to replace 
CT in a quantum-enabled future.

\begin{figure*}[htbp]
    \centerline{\includegraphics[width=0.8\linewidth]{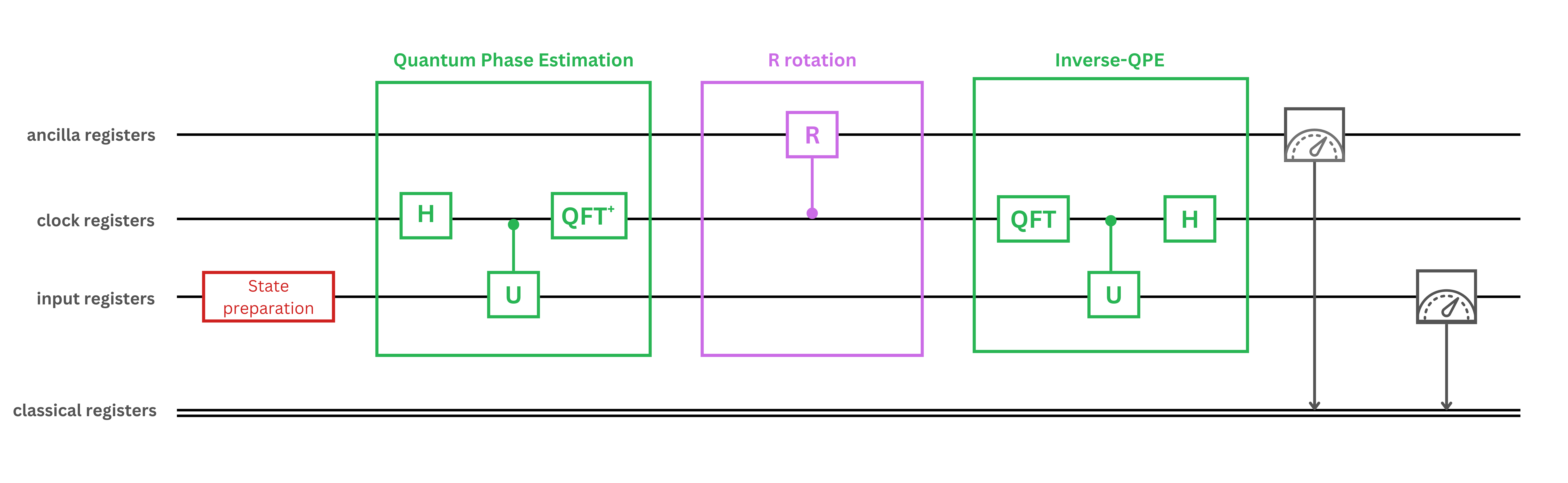}}
    \caption{The schematic of the HHL Algorithm shows the HHL algorithm working in 
    three stages: (a) Quantum Phase Estimation is used to encode the eigenvalues of 
    matrix $A$ into a clock register; (b) A controlled rotation is 
    applied to encode the inverse of these eigenvalues into an ancilla qubit; (c)
     Inverse QPE is performed, and postselection on the ancilla yields a quantum 
     state proportional to the solution vector $|x\rangle$ of the linear system $A|x\rangle = |b\rangle$.}
    \label{block}
\end{figure*}

\begin{figure*}[htbp]
    \centerline{\includegraphics[width=0.9\linewidth]{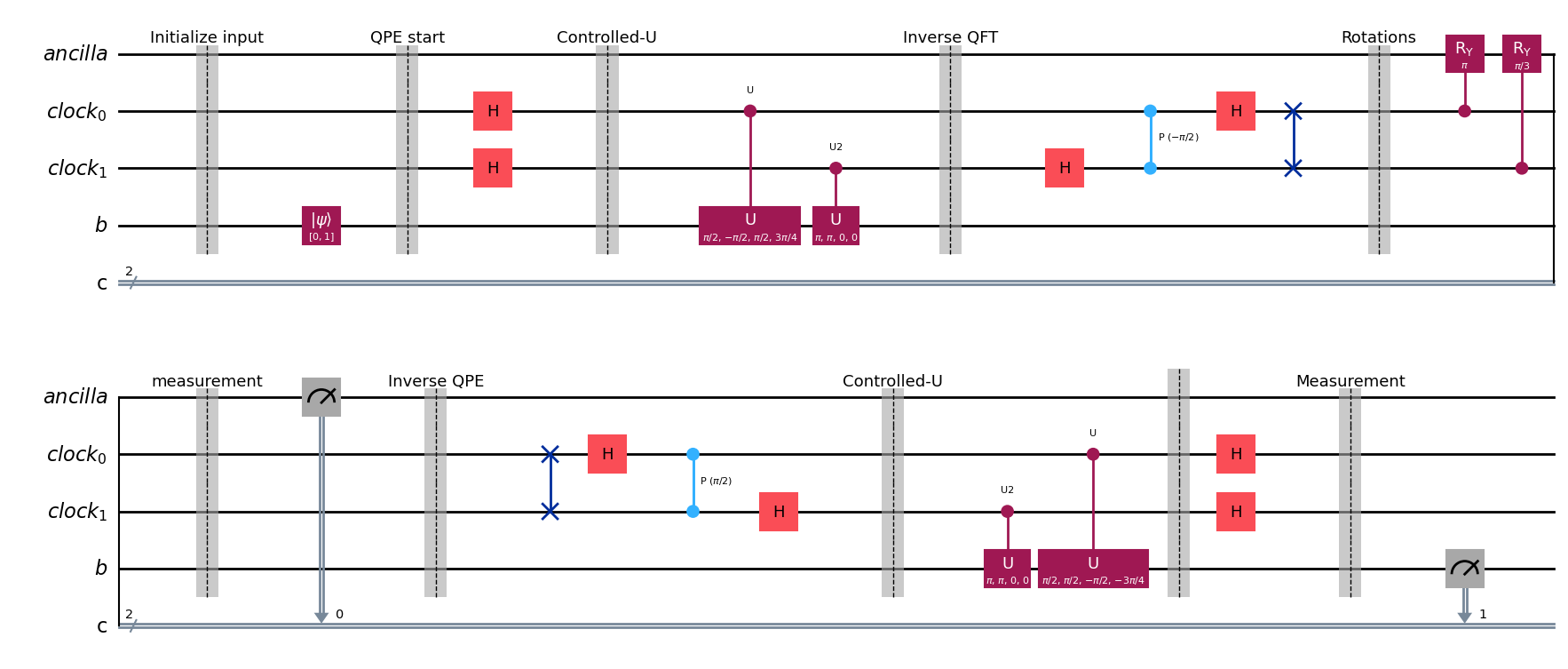}}
    \caption{Qiskit circuit representation of HHL Algorithm shows the main subroutines in it 
    including Quantum Phase Estimation, controlled rotations and inverse Quantum 
    Phase Estimation. Quantum Phase Estimation begins with a series of Hadamard gates on the clock 
    registers, followed by controlled-unitary operations ($U = e^{iAt}$) and 
    ends with an inverse Quantum Fourier Transform (inverse-QFT) composed of Hadamard gates, 
    controlled phase rotations, and SWAP gates. Controlled rotation consists of ancilla qubits 
    undergoing a controlled $R_y$ rotation conditioned on the eigenvalue 
    register to encode $1/\lambda$. Inverse QPE uncomputes the QPE solution, 
    leaving the solution encoded in the eigenvector register. 
    A measurement on the ancilla register postselects the successful outcome of the b register.}
    \label{HHL}
\end{figure*}

HHL consists of three main quantum subroutines as presented in Fig. \ref{block} and Fig. \ref{HHL}:

\begin{enumerate}
    \item \textbf{Quantum Phase Estimation (QPE):} The eigenvalues \( \lambda_j \) of \( A \) are encoded 
    into a quantum register using phase estimation on the unitary operator \( e^{iAt} \). The 
    input state \( |b\rangle \), expressed in the eigenbasis of \( A \), evolves into:
    \begin{equation}
    \sum_j \beta_j |v_j\rangle |\lambda_j\rangle
    \end{equation}

    \item \textbf{Controlled Rotation \cite{motto}:} An ancilla qubit is rotated based on \( 1/\lambda_j \). 
    This step encodes the inverse of the eigenvalues into amplitudes:
    \begin{equation}
    \sum_j \beta_j \frac{1}{\lambda_j} |v_j\rangle |1\rangle
    \end{equation}
    where success depends on measuring the ancilla in the \( |1\rangle \) state.

    \item \textbf{Inverse-QPE and Post-selection \cite{liu}:} Phase estimation is reversed (uncomputed), 
    and the system collapses to a quantum state approximating the normalized 
    solution \( |\tilde{x}\rangle \propto A^{-1}|b\rangle \). 
\end{enumerate}

\section{Comparative Analysis of ART and HHL for CT Reconstruction}

\subsection{Core Mathematical Foundation and Applicability}
Both ART and HHL algorithms fundamentally solve the system of linear equations related 
to \(A x = b\). This represents 
the essential mathematical model in computed tomography. In this context, \(x\) corresponds 
to the unknown voxel intensities of the medical image, \(b\) is the measured projection data, 
and \(A\) encodes the system geometry. ART is widely recognized for its capability in handling 
large, sparse matrices efficiently and performs robustly with noisy data that is commonly 
encountered in real-world CT scans.~\cite{Tang2012} Similarly, the quantum-based HHL algorithm is theoretically 
well-suited to large sparse systems. The condition for it is that 
the coefficient matrix is Hermitian ~\cite{harrow2009}.

\subsection{Input Format and Preprocessing}

ART operates directly with classical 
data obtained from projection measurements. The input vector \( b \) 
is straightforwardly loaded into classical computational memory. 
Each component \( b_i \) corresponds directly to an X-ray measurement and can be 
accessed individually or sequentially using standard computational resources. 
ART does not require special preprocessing or 
complex data structures making it highly flexible 
and practically efficient. \cite{ander1989}

\begin{figure*}[htbp]
    \centering
    \subfloat[Initialization of a 4-dimensional vector.]{%
      \includegraphics[width=0.35\columnwidth]{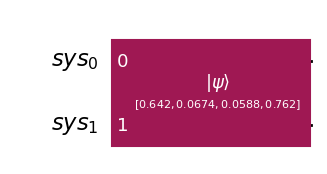}%
      \label{fig:subfig-0}%
    }\hfil
    \subfloat[Initialization of an 8-dimensional vector.]{%
      \includegraphics[width=0.5\columnwidth]{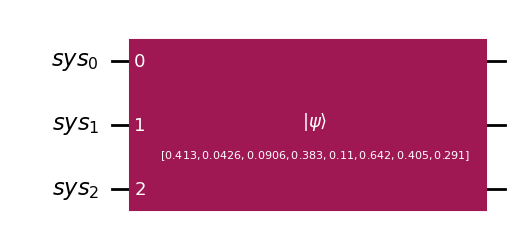}%
      \label{fig:subfig-1}%
    }\hfil
    \subfloat[Initialization of a 16-dimensional vector.]{%
      \includegraphics[width=0.7\columnwidth]{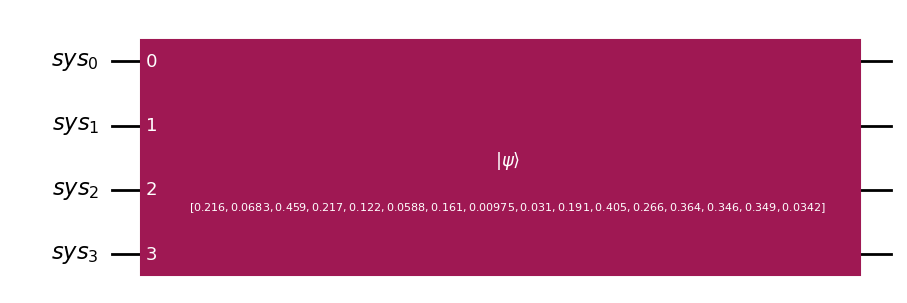}%
      \label{fig:subfig-2}%
    }
    \caption{The Qiskit-generated state-preparation circuits for different vector sizes
    shows how data is loaded into \(\log_2(n)\) qubits for an arbitrary vector. Only the 
    input qubits are shown here. 
    The number of clock qubits then 
    depend on the number of input or system qubits used and the required accuracy of the QPE 
    algorithm. Only one ancilla qubit is needed for all.}
    \label{fig:init}
\end{figure*}

In contrast, the HHL algorithm requires a fundamentally 
different way to input data. Rather than conventional arrays, HHL requires 
the vector \( b = (b_1, b_2, \dots, b_n) \) to be encoded directly into the 
amplitudes of quantum states. Specifically, HHL prepares the quantum state:
\begin{equation}
|b\rangle = \sum_{i=1}^{n} b_i |i\rangle
\end{equation}
where \( |b\rangle \) is a quantum state consisting of \(\log_2(n)\) qubits, with amplitudes 
encoding the classical input data \( b \). This fact is shown in Fig. \ref{fig:init} In theory, encoding vector \( b \) into quantum 
memory can be achieved by quantum RAM, capable of storing classical data values \( b_i \) and 
loading them simultaneously into quantum superposition. The efficiency of encoding and subsequent 
usage within the HHL algorithm is sensitive to the distribution of entries in \( b \). 
If the vector \( b \) contains a few elements \( b_i \) significantly larger than others, 
efficiently encoding and loading it into a quantum state becomes difficult. 
If quantum RAM is unavailable or impractical, 
quantum state preparation may rely on explicit classical preprocessing or quantum gate 
sequences. If preparation of the quantum state \( |b\rangle \) requires \( n^c \) computational 
steps for some constant \( c \) then the exponential speedup promised by HHL effectively disappears 
at the initial input preparation stage itself. ~\cite{aaronson2014} ~\cite{dervo2018}

\subsection{Runtime and Computational Complexity}
ART typically exhibits a worst-case 
computational complexity of $\mathcal{O}(n^3)$ particularly 
when large-scale iterative updates are involved. In practical 
implementations variants such as SART (Simultaneous ART) or regularized solvers reduce this 
burden. But ART still scales poorly with high-resolution volumetric data. The HHL 
algorithm offers a theoretical exponential speedup by solving 
linear systems in $\mathcal{O}(\log n)$ time under ideal conditions. 
According to Zaman, Morrell and Wong ~\cite{Zaman2023}, 
HHL accomplishes this speedup by leveraging quantum subroutines 
such as Quantum Phase Estimation (QPE) and Hamiltonian simulation which enable 
eigenvalue encoding and solution retrieval in the amplitude of quantum states. 

\subsection{Output Format and Interpretability}

In ART, the output format is straightforward and immediately interpretable. 
After reconstruction ART directly yields the solution vector \( x = (x_1, x_2, \dots, x_n) \), 
where each entry \( x_i \) represents a voxel intensity within the reconstructed medical image. 
ART operates entirely within classical computing paradigms. 
So, the reconstructed vector \( x \) can be easily visualized, interpreted, and utilized 
directly in clinical decision-making. Visualization tools, diagnostics, and analyses can 
immediately be applied to the output without additional complexity 
or significant post-processing. \cite{hanna2022}

The HHL algorithm does not 
directly yield the classical solution vector \( x \). Instead, it produces a quantum state:
\begin{equation}
|x\rangle = \sum_{i=1}^{n} x_i |i\rangle
\end{equation}

encoded within \(\log_2(n)\) qubits. The amplitudes of this quantum state approximate 
the entries of the solution vector \( x \). Thus, the algorithm's solution exists 
only implicitly within quantum amplitudes and not explicitly as classical data.
The user can only extract limited statistical information from this 
quantum state through quantum measurement. For example, measurements can 
identify positions of unusually large entries in the solution vector or estimate 
inner products \( \langle x | z \rangle \) with predetermined vectors \( z \). However, 
retrieving the precise numerical value of a specific entry \( x_i \) generally requires 
repeated executions of the HHL algorithm for \( n \) repetitions to achieve 
reliable estimates. This requirement for repetition effectively reduces the exponential 
computational advantage initially promised by the quantum algorithm. \cite{aaronson2014}

\begin{figure}[htbp]
    \centerline{\includegraphics[width=0.7\linewidth]{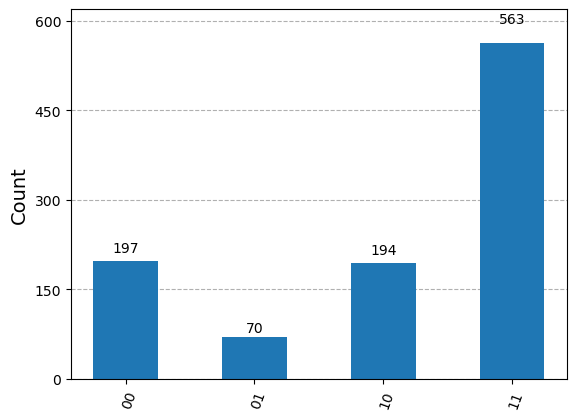}}
    \caption{The results of simulation of the circuit presented in Fig. \ref{HHL} shows the 
    variation of the results with the number of shots. The 
    two classical bits represent outcomes from measuring the ancilla qubit (bit 0) and 
    the solution qubit (bit 1). A result of '11' indicates that both the 
    ancilla and solution qubits collapsed to the \(|1\rangle\) state upon measurement. 
    Specifically, the ancilla qubit being in \(|1\rangle\) confirms successful 
    application of the controlled rotation, while the 
    solution qubit measured in \( |1\rangle \) reflects the corresponding 
    component of the quantum solution vector x. 
    Out of 1024 total shots, 563 shots resulted in '11', while the remaining 
    shots yielded other outcomes. This distribution highlights a fundamental aspect 
    of quantum computing, results are inherently probabilistic due to quantum 
    superposition and measurement collapse. Even in ideal conditions, 
    quantum states do not deterministically yield a single outcome but instead provide a 
    probability distribution over possible states, reflecting the amplitudes of 
    the quantum state prior to measurement.}
    \label{result-noiseless}
\end{figure}
\subsection{Noise Tolerance and Robustness}
ART is inherently robust to noise due to its iterative nature. In computed tomography noise arises 
from various sources including photon statistics and electronic fluctuations. 
ART addresses this by iteratively refining the solution vector \( x \) to minimize the 
discrepancy between the measured projections and the projections calculated from the 
current estimate of \( x \). This process allows ART to reduce the 
effects of noise in the data. ART can also incorporate prior knowledge and 
regularization techniques to enhance noise tolerance. Constraints such 
as non-negativity or smoothness can be applied to the solution to improve the quality of the 
reconstructed image in the presence of noise.\cite{kera2016} However, it's important 
to note that excessive noise can still impact the convergence and accuracy of 
ART. This requires careful selection of relaxation parameters and stopping criteria. \cite{green1974} 

The HHL algorithm operates within the quantum computing paradigm where noise is generated 
differently compared to classical systems. Quantum noise such as decoherence and gate 
errors poses significant challenges to the practical implementation of HHL. 
The algorithm's reliance on QPE and controlled rotations makes 
it particularly sensitive to such noise as these operations require high fidelity to 
maintain the integrity of the quantum state. 
Recent studies have explored the resilience of HHL to noise on Noisy 
Intermediate-Scale Quantum (NISQ)\cite{khanal2023} devices. Findings indicate that current noise 
mitigation techniques are insufficient to fully counteract the 
impact of quantum noise on HHL's performance. 
The algorithm's sensitivity to noise requires the development of 
fault-tolerant quantum computing architectures and advanced error correction methods 
to realize its theoretical advantages in practical applications. \cite{bertrand2024} 

\subsection{Hardware Readiness and Implementation}
ART is well-established within classical computational frameworks. Its implementation 
relies on readily available computing hardware. GPUs  
have significantly accelerated ART by parallelizing matrix-vector computations, 
iterative updates, and back-projection operations. 
Modern computing infrastructures and widely accessible software libraries 
ensure that ART is both practical and scalable for routine clinical deployment. This 
is true even in high-resolution, large-scale CT imaging scenarios.  
ART faces minimal hardware limitations and has benefitted from decades of 
optimization in computational resources and technological advancements. \cite{hanna2022}

In contrast, quantum computing hardware is currently at the early stages of development.\cite{bravi2022} 
This can 
be seen in the significant limitations in qubit count, coherence times, and error rates. 
Real-world CT reconstruction tasks typically involve large-scale datasets with millions of 
variables and so require quantum computers with thousands to millions of logical qubits. 
Current quantum hardware is limited to around a hundred noisy physical qubits. This  
makes practical-scale CT reconstruction unfeasible at present. 
The HHL algorithm also requires precise quantum operations which demand highly 
reliable quantum gates and extended coherence times. \cite{tripathai}
While significant research is ongoing in quantum hardware development and 
quantum error correction, substantial technological breakthroughs remain 
necessary before practical medical imaging applications can become viable.

\subsection{Flexibility and Customizability}
ART is highly flexible and customizable. This makes it well-suited for diverse 
clinical and research applications. ART allows straightforward integration of 
constraints, and regularization techniques to enhance reconstruction 
quality. Parameters such as relaxation factors, stopping criteria, and regularization terms 
can be easily tuned to optimize performance for different imaging conditions and clinical 
scenarios. ART can also accommodate modifications tailored specifically for 
parse or noisy datasets which enables practitioners to adjust the reconstruction 
algorithm according to their specific clinical or research objectives. 
This flexibility ensures ART remains adaptable and effective across a wide range of CT imaging contexts. \cite{fager}

The flexibility and customizability of the HHL algorithm are currently limited 
by the underlying quantum hardware and algorithmic constraints. HHL 
could incorporate custom-designed quantum subroutines and tailored unitary 
transformations to handle specific linear system characteristics. However, practical 
limitations imposed by current quantum hardware significantly restrict its adaptability. \cite{baskaran2022}
The requirement that the matrix \( A \) be sparse, Hermitian, and well-conditioned places 
strict constraints on the types of problems that can be effectively solved by HHL. 
Customizing HHL for different problems also often requires developing specialized 
quantum circuits and advanced quantum state encoding methods which are currently challenging due 
to hardware constraints and quantum noise. \cite{dari} 

\subsection{Future Potential and Research Directions}
The future potential of ART revolves around incremental 
improvements rather than transformative breakthroughs. Research directions include 
further optimization of computational efficiency, enhancement of noise robustness, 
and refinement of regularization strategies. \cite{donato2022}  Advanced machine learning methods, such 
as deep neural networks integrated with ART \cite{mcleavy2021}, represent 
promising directions. 

While, the HHL algorithm presents transformative future potential driven by ongoing 
advancements in quantum computing. Key research directions include the development of 
fault-tolerant quantum computers capable of handling 
large-scale clinical datasets and mitigating quantum noise. 
Another critical direction involves creating efficient quantum data encoding and 
decoding techniques to bridge the gap between classical data acquisition and quantum processing. 
Hybrid quantum-classical frameworks also represent a promising research area. 
Such frameworks could leverage quantum algorithms like HHL for specific computational bottlenecks  
complemented by classical methods like ART \cite{Yan2024} to handle more conventional tasks. 
Quantum-assisted feature extraction, quantum-enhanced regularization, and 
quantum-driven data compression also represent potential avenues through which HHL could 
enhance medical imaging significantly. \cite{morales2024}

\section{Implementing the HHL algorithm for a small-scale image reconstruction}

\begin{figure}[htbp]
    \centerline{\includegraphics[width=0.55\linewidth]{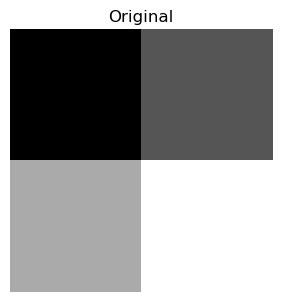}}
    \caption{The $2\times2$ phantom 
    used as the test image shows four distinct contrast 
    values to create a clear test for evaluating the accuracy of the reconstruction algorithms.}
    \label{phantom}
\end{figure}

To examine how the promise of HHL translates into the CT domain, we built 
an end-to-end miniature experiment that places a classical ART solver and an HHL-based solver 
side-by-side on the smallest meaningful case \cite{githubrepo}: reconstructing a $2\times2$ 
phantom image \cite{wenger}. A phantom is a synthetic image with known ground truth. The 
phantom image used in this test is shown in Fig. \ref{phantom} 
The problem was run on the IBM Qiskit \cite{java} \cite{qiskit} \cite{qiskitdocs} environment in 
Python using a noiseless
AerSimulator \cite{tan} backend. 

\begin{figure}[htbp]
    \centerline{\includegraphics[width=0.6\linewidth]{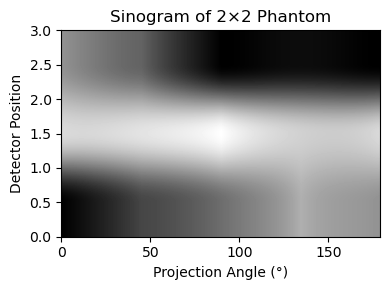}}
    \caption{The sinogram is the 
    raw dataset from which algorithms back-project and reconstruct cross-sectional CT images. 
    The sinogram looks like what it does because each point in the 
    object traces a sinusoidal path in the sinogram because, as the scanner rotates, 
    the position of that feature's projection shifts in a predictable, periodic way.}
    \label{sino}
\end{figure}

In CT each pixel is never measured directly. Detectors record line-integral projections.  
Combined projection data collected at rotation angles are arranged into a 
two-dimensional array called a sinogram \cite{zakar}.  Each row records detector readings for one angle, 
so points in the object trace sinusoidal paths across the array. The sinogram is the 
raw dataset from which algorithms back-project and reconstruct cross-sectional CT images. The 
sinogram for the test is represented in Fig. \ref{sino}
An established ART algorithm is used to reconstruct the images from this dataset, which is quite successful.

The implementation of the HHL algorithm was not so straightforward.
A direct HHL application requires $A$ to be Hermitian and well-conditioned.  
Neither is true, so we solve the normal equations  
\[
(\,A^{\!\mathsf T}A+I)\,\mathbf{x}=A^{\!\mathsf T}\mathbf{b},
\]
which are Hermitian, positive-definite, and exhibit required eigenvalues.  
Adding $I$ keeps eigenvalues away from 0. This improves robustness at the cost of a slight downward bias 
and removes the divide by 0 error we may face in HHL implementation (we encode $1/\lambda$ in the 
controlled rotation gates).
The text block below shows the input matrices and how it is initialized in the HHL algorithm.

\begin{lstlisting}
    Projection matrix A (4x4):
    [[1. 0. 1. 0.]
    [0. 1. 0. 1.]
    [1. 1. 0. 0.]
    [0. 0. 1. 1.]]

    Transpose A^T:
    [[1. 0. 1. 0.]
    [0. 1. 1. 0.]
    [1. 0. 0. 1.]
    [0. 1. 0. 1.]]

    b (4x1):
    [[4.]
    [6.]
    [3.]
    [7.]]

    Hermitian matrix A_herm = A4^T A4 + I:
    [[3. 1. 1. 0.]
    [1. 3. 0. 1.]
    [1. 0. 3. 1.]
    [0. 1. 1. 3.]]

    b_herm = A^T b:
    [[ 7.]
    [ 9.]
    [11.]
    [13.]]

    Normalized vector b_norm (used in state initialize):
    [[0.34156503]
    [0.43915503]
    [0.53674504]
    [0.63433505]]


\end{lstlisting}

\begin{figure*}[htbp]
    \centerline{\includegraphics[width=0.9\linewidth]{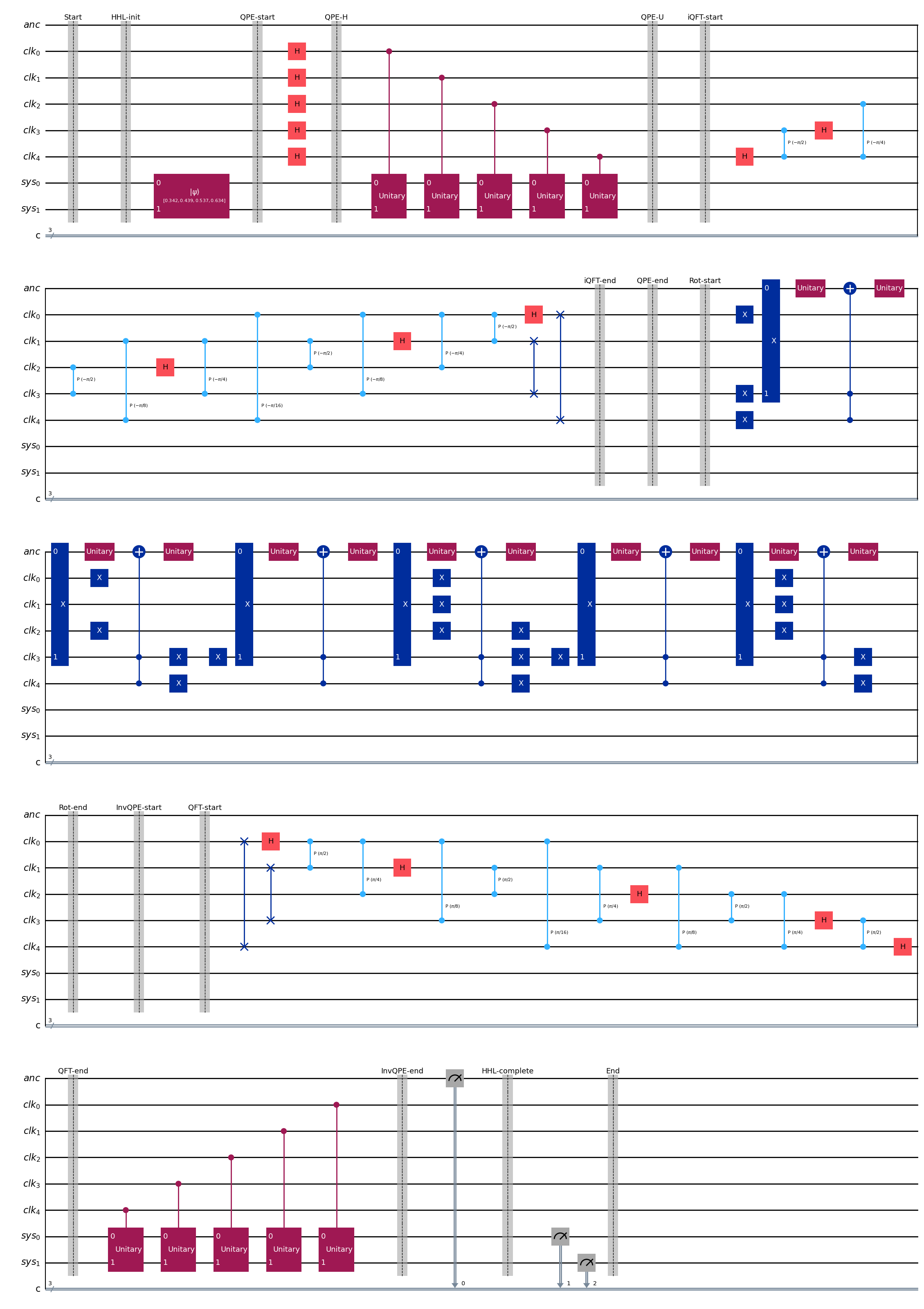}}
    \caption{Quantum circuit implementation of the small-scale HHL algorithm in Qiskit shows 
    that the circuit comprises an ancilla qubit for eigenvalue inversion, five 
    phase-estimation “clock” qubits, and two system or input qubits encoding the four-pixel 
    image vector. The controlled rotation gates in the HHL algorithm are complex to implement 
    because they require encoding the inverse of eigenvalues ($1/\lambda$). 
    This is challenging due to precision limits, handling small eigenvalues, and 
    designing conditional rotations based on quantum-encoded values.
    }
    \label{hhl-toy}
\end{figure*}

HHL's Quantum Phase Estimation (QPE) needs controlled unitaries \cite{tana} $e^{\mathrm{i}At}$.  
For the matrix the exponential is computed numerically, and embedded in controlled gates.  

The final circuit implementation of HHL contains one ancilla qubit for eigenvalue reciprocals,
three ''clock'' qubits for QPE, and two system qubits for the four-pixel state vector. This is 
shown by qiskit's circuit representation in fig. \ref{hhl-toy}.

The algorithm then followed the following steps:
\begin{enumerate}
    \item initializes the system qubits in the normalised vector,
    \item performs QPE to entangle eigen-phases,
    \item conditionally rotates the ancilla to encode $1/\lambda_j$,
    \item reverses QPE to disentangle phases,
    \item measures the ancilla, post-selecting successful inversions.
\end{enumerate}
A state-vector simulation \cite{jama} then extracts the unmeasured amplitudes of the system register.
The image is then reconstructed using the amplitudes of the system register.

\begin{figure*}[htbp]
    \centerline{\includegraphics[width=0.6\linewidth]{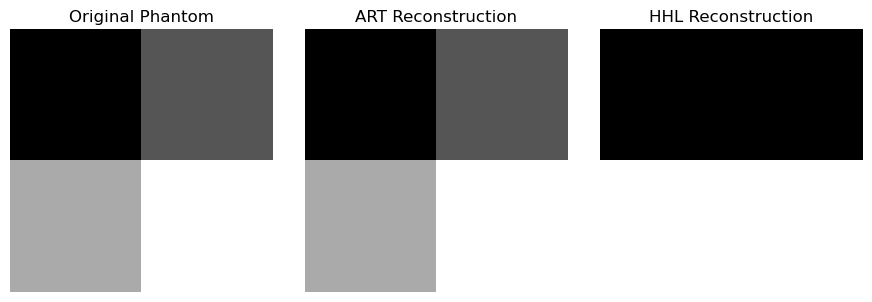}}
    \caption{The comparison of image reconstruction of ART vs HHL shows almost what we expected it to be.
    ART is an established and highly accurate reconstruction algorithm and thus reconstructed image is 
    exactly as the original image. The HHL quantum algorithm is noisy and thus the darker parts 
    in the top look the same and the lighter parts on the bottom are the same. 
    This is the highest accuracy we could get with a high number of 
    clock qubits implemented as more clock qubits equal accurate QPE routine.}
    \label{result}
\end{figure*}

A comparison of the original image, its ART reconstruction and the HHL reconstruction 
is shown in fig. \ref{result}.
The reconstructed pattern is correct for ART and almost correct for HHL. However, this level of 
inaccuracy is not suitable for medical application as CT image reconstruction requires a 
very high level of precision that is not possible with modern day quantum hardware.

\section{Key Challenges Identified in the Practical Implementation of HHL}

After a small-scale implementation of the HHL, the following key challenges have been identified:

\subsection{Quantum Data Encoding Complexity}
Unlike ART, which operates directly on classical projection data, 
the HHL workflow first demands that the linear system be mapped to a quantum-compatible form.  
The coefficient matrix \(A\) must be embedded in a Hermitian operator by forming the 
normal equations \((A^{\mathsf T}A+I)\mathbf{x}=A^{\mathsf T}\mathbf{b}\).  Simultaneously, 
the right-hand-side 
vector must be amplitude-encoded on the system qubits as the circuit's initial state.  Engineering 
these controlled unitaries and preparing the input state generally requires \(\mathcal{O}(n^{c})\) 
gate operations for realistic problem sizes, a scaling overhead that can erode HHL's theoretical 
exponential speed-up when applied to large datasets. \cite{Zaman2023}

\subsection{Quantum Output Interpretability}
HHL outputs solutions implicitly encoded in quantum states and each 
solution entry is represented as quantum amplitudes. Direct 
extraction of specific voxel intensities requires multiple repeated 
quantum measurements and algorithm executions. To accurately retrieve 
individual solution values approximately \( n \) repetitions of the algorithm are necessary which  
significantly undermines the exponential advantage promised by quantum algorithms for classical 
data retrieval.
In our toy \(2\times2\) study we sidestepped the issue by using a
state-vector simulator, which grants direct access to the full wavefunction in a single run. Such 
global readout is infeasible on current quantum hardware.

\subsection{Quantum Noise and Robustness}
The sensitivity of the HHL algorithm to quantum noise presents a 
substantial challenge. Quantum computing currently suffers from decoherence 
and gate operation errors which severely affects the performance of quantum subroutines 
needed for HHL. The reliance of HHL on precise quantum state 
manipulation magnifies its vulnerability to noise. This necessitates fault-tolerant quantum 
architectures and advanced error-correction techniques not yet available on practical scales.

\subsection{Hardware Limitations}
Practical CT imaging tasks require quantum processors 
capable of handling millions of data points which translates to millions of 
logical qubits. Current quantum hardware is restricted to roughly a hundred 
noisy physical qubits \cite{ichi} with limited coherence times. Moreover, the quantum operations 
integral to HHL (like controlled rotations and QPE) require high fidelity and low error rates which are 
currently unattainable on existing quantum hardware platforms.

\subsection{Algorithmic Flexibility Constraints}
HHL's practical adaptability is currently limited due to stringent 
algorithmic constraints. The requirement that the coefficient matrix \( A \) must be sparse, 
Hermitian, and well-conditioned restricts the algorithm's applicability to a relatively 
narrow range of CT scenarios. Customizing HHL for varied clinical applications involves 
specialized quantum circuits and complex encoding strategies which remain difficult under 
present hardware and noise constraints.

\section{Potential Roles and Future Integration of HHL in Computed Tomography}

Addressing the challenges through continued research and technological advancement is essential to 
realize the transformative promise of quantum computing in medical imaging. 
The HHL algorithm holds 
substantial promise for enhancing ART 
in computed tomography. While immediate full replacement of 
ART is improbable, targeted integration and specialized applications of HHL offer viable 
pathways for innovation and improvement within medical imaging workflows. The following 
section explores specific potential roles and integration scenarios for HHL in computed tomography despite 
the limitations.

\subsection{Hybrid Quantum-Classical Reconstruction Models}
Hybrid quantum-classical reconstruction models have emerged as a promising 
approach to integrating quantum computing capabilities into classical computational workflows.
These are particularly relevant for computationally demanding tasks. 
These hybrid frameworks leverage the advantages of both classical and quantum 
computation and addresses the limitations imposed by current quantum hardware constraints.

Yalovetzky, Minssen, Herman, and Pistoia \cite{yalo2024} introduced the Hybrid HHL++ algorithm, 
an advancement of the HHL quantum algorithm specifically 
tailored for execution on NISQ hardware. Hybrid HHL++ 
is designed to efficiently solve linear systems by combining classical preprocessing and 
optimization steps with quantum processing. Classical methods are utilized to determine 
optimal scaling factors and to perform circuit compression. This reduces the quantum 
resource overhead significantly. By doing this the hybrid approach mitigates challenges 
associated with qubit limitations, decoherence, and operational noise inherent in current 
quantum systems. Consequently, Hybrid HHL++ enhances the practical feasibility of quantum 
linear solvers, making it possible to address larger-scale linear systems than would be 
achievable using quantum computing alone. 

Building upon a similar hybrid computational paradigm, Ye et al. \cite{ye2024hybrid} presented a 
quantum-classical framework designed for computational fluid dynamics (CFD) \cite{ander1995} applications. 
CFD may differ in application context from CT but the underlying computational methodology 
shares critical similarities including the requirement to efficiently solve large-scale 
linear systems. Their approach utilizes quantum algorithms to tackle specific 
computational subproblems within the classical iterative reconstruction pipeline. 
In this model quantum algorithms, including variants of HHL, solve key sub-tasks 
involving linear algebra operations that are computationally intensive for classical solvers. 
Classical algorithms subsequently handle iterative refinement, convergence control, and 
result interpretation. This hybridization approach capitalizes on quantum 
advantages and also ensures robustness and practicality by combining classical 
and quantum strengths.

Integrating these insights into computed tomography reconstruction indicates a viable and beneficial 
direction. Hybrid quantum-classical methods could significantly improve reconstruction 
performance by selectively employing quantum algorithms to rapidly solve critical linear
systems derived from projection data. Classical components of the 
reconstruction pipeline would continue managing data preprocessing, 
noise handling, iterative refinement, and image interpretation. 
This complementary integration can potentially yield faster convergence, 
higher reconstruction accuracy, and more efficient handling of large 
datasets common in medical imaging applications.

\subsection{Efficient Representation of Classical Data in Quantum Computation}

Encoding classical data into quantum states efficiently is 
essential for the practical implementation of HHL. As discussed by Mitarai, Kitagawa, 
and Fujii \cite{mitarai2019quantum}, 
classical data can be encoded in quantum systems via two approaches: 
analog and digital encoding. In analog encoding, data is stored as amplitude 
coefficients of basis states that allow for compact representation of large vectors using fewer qubits. 
This method is used in HHL to encode the input vector $\vec{b}$ and enables exponential compression.
Digital encoding stores classical data as binary strings across quantum registers. 
This format is more suitable for performing arithmetic operations and is widely 
used in quantum optimization and machine learning applications. Algorithms 
often require transformation between these encoding types to leverage both 
advantages.

They propose two key techniques for this purpose. Quantum Digital-to-Analog 
Conversion (QDAC) \cite{mitarai2019quantum} transforms a 
digitally encoded quantum state into an analog format using 
controlled rotations. This approach is probabilistic and can be amplified through amplitude 
amplification. Quantum Analog-to-Digital Conversion (QADC) converts amplitude information from 
an analog state into digital bitstrings using swap tests, phase estimation, and quantum arithmetic. 
This process is deterministic and enables the digital extraction of the real, imaginary, or absolute 
values of amplitudes. 
These conversions are crucial for enabling nonlinear transformations on quantum states and  
are necessary for advanced quantum algorithms and quantum machine learning. They 
also enable flexible preprocessing of classical input data for use in quantum linear solvers 
like HHL. Such encoding allows efficient loading and manipulation of 
high-dimensional projection data within quantum circuits and helps in improving integration between quantum 
and classical workflows.

\subsection{Enhancement through Quantum Error Correction (QEC) \cite{cory}}
The deployment of HHL on NISQ 
hardware is hindered by sensitivity to quantum noise. QPE 
and Hamiltonian simulation are particularly vulnerable to decoherence 
and gate imperfections.
In their detailed analysis, Phillips \cite{philips2024} examine how eigenvalue 
approximation errors 
introduced during QPE propagate through the algorithm. Their results demonstrate that 
small errors in estimating eigenvalues \(\lambda_j\) lead to significant entanglement 
between the phase and flag registers and reduce the fidelity of the post-selected solution 
state. The deviation from the ideal output state is bounded by \(\mathcal{O}(\kappa / t_0)\), 
where \(\kappa\) is the condition number and \(t_0\) is the simulation duration. This 
relationship underscores the precision-resource tradeoff in practical implementations.

QEC provides a robust framework to overcome these limitations. 
By encoding logical qubits in error-resilient subspaces, QEC can protect the HHL circuit from 
bit-flip, phase-flip, and more general errors. When combined with fault-tolerant quantum 
gates this enhances the reliability of deeper circuits required for fine-grained 
eigenvalue resolution. QEC can also enable larger clock register sizes \(T\) and longer 
simulation times \(t_0\). Both of them are crucial for reducing approximation error in the 
QPE step. As quantum hardware evolves, the integration of QEC into hybrid quantum-classical 
HHL pipelines will likely be essential for realizing the algorithm's full potential in scientific 
computing and imaging applications.

\subsection{Rapid Advancement in Quantum Computation}
Quantum computing has rapidly evolved from theoretical constructs into an 
experimental and increasingly practical discipline and pushing the boundaries of what 
is computationally possible. As outlined by Memon, Al Ahmad, and Pecht \cite{memon2024}, this field is 
undergoing transformative advancements across hardware platforms, enabling scientists and 
engineers to inch closer toward quantum advantage—solving problems intractable for classical 
computers.
Key breakthroughs are being made across various quantum hardware platforms. 
Superconducting qubits which are one of the most mature architectures have seen 
improvements in scalability, error mitigation, and cryogenic infrastructure. 
Companies like IBM \cite{qiskit} and Google \cite{abu} have demonstrated multi-qubit systems that maintain 
coherence for longer durations enhancing circuit depth and algorithmic fidelity.
Trapped-ion qubits \cite{srinivas} offer another promising path. They boast exceptionally high-fidelity 
gates and long coherence times. Recent innovations have minimized the complexity of 
their control systems, improving their practicality for larger-scale implementations. 
Photonic qubits are also being 
explored for room-temperature quantum computing. \cite{bravi2022}
The convergence of AI, quantum simulation, and 
quantum machine learning \cite{khanal2024} is shaping the development of new applications in this
sector. 
Such advancements in quantum computation are essential for 
implementing algorithms like HHL in medical imaging, a domain that is 
both computationally intensive and demands high accuracy with minimal tolerance for error.

\section{Educational Insights and Recommendations}
By implementing the complete HHL algorithm for a small scale CT reconstruction, 
we turned abstract quantum-computing ideas into concrete tasks.  We wrote, 
debugged, and benchmarked every subroutine ourselves. This allowed us to shift 
the basic concepts of quantum computing, specifically the HHL algorithm, from theory to experience.  
We also learned problem solving by learning to dive into primary literature 
when documentation failed. This also allowed us to judge if the advertised quantum speed-up is 
neutralized by data-loading overhead.

For an introductory quantum computing course, introducing the assignment early in the semester 
seems to be highly beneficial. This approach allows students to engage with 
fundamental concepts in the context of their own meaningful, practical project. This can also help to  
reinforce ideas as they are learned. It also provides ample time for iteration, reflection, and 
deeper understanding.
This also keeps students motivated by connecting theoretical knowledge with hands-on 
application throughout the course. For instance, our own project choice 
enabled us to focus specifically on essential quantum computing concepts such as 
Quantum Phase Estimation (QPE) and the Quantum Fourier Transform (QFT) which 
are critical subroutines within the HHL algorithm. 
Qiskit remains the most accessible platform
because it pairs a Python interface with exact state-vector
simulation and a growing set of pedagogical notebooks. 

We noted that the assignment works best when presented as a narrative rather than
a checklist. Framing the project as “using quantum algorithm for CT scans” instantly 
motivates the students to push through any hardships
that follows. Providing a minimal but functioning skeleton
code lowers the activation energy. One of the main hurdles 
we faced was a lack of existing code, not even in a small-scale. Providing a basic 
code allows students to spend their cognitive
capital on conceptual hurdles.

We recommend incorporating a conference-style paper writing component too into
project-based assignments, as in our case. 
For graduate students, it provides structured practice in communicating complex technical ideas 
clearly in formal academic applications. 
It also introduces them to peer-reviewed formatting standards, citation practices, and the process of 
preparing work for submission to a real conference. For undergraduate students, the experience 
serves as an introduction to academic research and scientific writing. 
It fosters ownership of their learning and familiarizes them with the expectations of scholarly 
communication. In our experience, we found that the 
writing component not only reinforced the conceptual 
understanding but 
also deepened our appreciation of the broader implications and relevance of the work.

We also emphasize the need for a good interdisciplinary mentorship in project-based 
assignments that span multiple domains. In our project, since the topic bridged two 
different disciplines, support from faculty across both domains was absolutely necessary. 
Beyond technical instruction, mentors from both fields should ideally provide support with 
academic writing, especially regarding how to adapt technical content for a research audience. 
This dual-mentorship model not only enhances student outcomes but also prepares them 
for the collaborative, interdisciplinary nature of real-world scientific research.

In sum, the assignment's blend of project implementation,
and technical writing activates a range of skills that
traditional lecture-based courses rarely address.
Deployed at the right moment in the semester, with scaffolding that
balances guidance and autonomy, it can become a signature learning
experience that anchors an introductory quantum computing
curriculum.

\section{Conclusion}
This paper highlights the value of project-based learning (PBL) as an 
effective approach for teaching quantum computing concepts. 
By engaging with a real-world problem, we 
were able to move beyond theoretical understanding and actively explore how quantum algorithms 
perform in practice. The process of researching, implementing, and critically evaluating a quantum 
solution within a familiar context fostered deeper conceptual understanding, 
problem-solving skills, and technical communication abilities.

The assignment, using HHL for image reconstruction, demonstrated how PBL encourages 
learners to confront authentic challenges, from 
handling quantum data encoding to interpreting algorithmic limitations. More 
importantly, it provided a framework where students could develop critical thinking by 
assessing when quantum approaches are advantageous and when classical methods remain superior. 
We learned that ART remains a robust, practical method for CT reconstruction. However, quantum computing,
especially the HHL algorithm, offers transformative potential through faster 
linear system solutions. Despite current hardware and implementation challenges, 
hybrid quantum-classical approaches may soon enhance CT pipelines. 
As quantum technology matures it promises to advance medical imaging's computational foundations.

Our experience suggests that integrating structured PBL modules into introductory quantum 
computing courses can transform passive learning into active learning. By framing assignments 
around interdisciplinary applications educators 
can provide context-driven education. This enhances engagement and retention among students. Moreover, 
tasks that combine project implementation, and technical writing cultivate a 
broad skill set rarely achieved through traditional lecture formats.
Additionally, we recommend that educators consider incorporating a formal writing 
and submission component alongside interdisciplinary mentorship, as these elements 
greatly enhance students' technical communication skills and expose them to authentic research practices.

In conclusion, this project serves as a model for how PBL can be leveraged to teach quantum 
computing more effectively. We recommend that educators adopt similar approaches, where students 
are challenged to apply quantum principles to real-world problems, encouraging both technical 
mastery and critical evaluation.

\end{document}